\documentclass{article}

\usepackage[utf8]{inputenc}
\usepackage[margin=1in]{geometry}
\usepackage[numbers]{natbib}
\usepackage{graphicx}
\usepackage{amsmath}
\usepackage{amssymb}
\usepackage{amsthm}

\newcommand{\comment}[1]{}

\newcommand{\ch}[1]{ \{#1_t\}_{t \geq 0} }
\newcommand{\X}{\mathcal{X}}
\newcommand{\B}{\mathcal{B}}

\newtheorem{Theorem}{Theorem}
\newtheorem{Proposition}{Proposition}
\newtheorem{Assumption}{Assumption}

\title{Geometric ergodicity of the Random Walk Metropolis with position-dependent proposal covariance}

\author{
Samuel Livingstone
\\ 
\small \textit{Department of Statistical Science, University College London, UK}
\\
\small \texttt{samuel.livingstone@ucl.ac.uk}
}

\date{}

\begin{document}

\maketitle

\abstract{We consider a Metropolis--Hastings method with proposal $\mathcal{N}(x, hG(x)^{-1})$, where $x$ is the current state, and study its ergodicity properties. We show that suitable choices of $G(x)$ can change these compared to the Random Walk Metropolis case $\mathcal{N}(x, h\Sigma)$, either for better or worse. We find that if the proposal variance is allowed to grow unboundedly in the tails of the distribution then geometric ergodicity can be established when the target distribution for the algorithm has tails that are heavier than exponential, but that the growth rate must be carefully controlled to prevent the rejection rate approaching unity. We also illustrate that a judicious choice of $G(x)$ can result in a geometrically ergodic chain when probability concentrates on an ever narrower ridge in the tails, something that is not true for the Random Walk Metropolis.
}

\section{Introduction}
Markov chain Monte Carlo (MCMC) methods are techniques for estimating expectations with respect to some distribution $\pi(\cdot)$, which need not be normalised.  This is done by sampling a Markov chain which has limiting distribution $\pi(\cdot)$, and computing empirical averages.  A popular form of MCMC is the Metropolis--Hastings algorithm \cite{metropolis1953equation,hastings1970monte}, where at each time step a `proposed' move is drawn from some candidate distribution, and then accepted with some probability, otherwise the chain stays at the current point.  Interest lies in finding choices of candidate distribution that will produce sensible estimators for expectations with respect to $\pi(\cdot)$.

The quality of these estimators can be assessed in many different ways, but a common approach is to understand conditions on $\pi(\cdot)$ that will result in a chain which converges to its limiting distribution at a \emph{geometric} rate.  If such a rate can be established, then a Central Limit Theorem will exist for expectations of functionals with finite second absolute moment under $\pi(\cdot)$ if the chain is reversible.

A simple yet often effective choice is a symmetric candidate distribution centred at the current point in the chain (with a fixed variance), resulting in the \emph{Random Walk Metropolis} (RWM) (e.g. \cite{sherlock2010random}).  The convergence properties of a chain produced by the RWM are well-studied.  In one dimension, essentially convergence is geometric if $\pi(x)$ decays at an exponential or faster rate in the tails \cite{mengersen1996rates}, while in higher dimensions an additional curvature condition is required \cite{roberts1996geometric}.  Slower rates of convergence have also been established in the case of heavier tails \cite{jarner2007convergence}.

Recently, some MCMC methods have been proposed which generalise the RWM, whereby proposals are still centred at the current point $x$ and symmetric, but the variance changes with $x$ \cite{roberts2009examples,roberts2002langevin,sejdinovic2014kernel,andrieu2008tutorial,
craiu2009learn}.  An extension to infinite-dimensional Hilbert spaces is also suggested in \cite{rudolf2015generalization}.  The motivation is that the chain can become more `local', perhaps making larger jumps when out in the tails, or mimicking the local dependence structure of $\pi(\cdot)$ to propose more intelligent moves.  Designing MCMC methods of this nature is particularly relevant for modern Bayesian inference problems, where posterior distributions are often high dimensional and exhibit nonlinear correlations \cite{girolami2011riemann}.  We term this approach the \emph{Position-dependent Random Walk Metropolis} (PDRWM), although technically this is a misnomer, since proposals are no longer random walks.  Other choices of candidate distribution designed with distributions that exhibit nonlinear correlations were introduced in \cite{girolami2011riemann}.  Although powerful, these require derivative information for $\log\pi(x)$, something which can be unavailable in modern inference problems (e.g. \cite{brooks2011handbook}).  We note that no such information is required for the PDRWM, as evidenced by the particular cases suggested in \cite{roberts2009examples,roberts2002langevin,sejdinovic2014kernel,andrieu2008tutorial,craiu2009learn}.  However, there are relations between the approaches, to the extent that understanding how the properties of the PDRWM differ from the standard RWM should also aid understanding of the methods introduced in \cite{girolami2011riemann}.

In this article we consider the convergence rate of a Markov chain generated by the PDRWM to its limiting distribution.  Our main interest lies in whether this generalisation can change these \emph{ergodicity} properties compared to the standard RWM with fixed covariance.  We focus on the case in which the candidate distribution is Gaussian, and illustrate that such changes can occur in several different ways, either for better or worse. Our aim is not to give a complete characterisation of the approach, but rather to illustrate the possibilities through carefully chosen examples, which are known to be indicative of more general behaviour.

In Section \ref{sec:markov} necessary concepts about Markov chains are briefly reviewed, before the PDRWM is introduced in Section \ref{sec:pdrwm}.  Some results in the one-dimensional case are given in Section \ref{sec:1d}, before a model higher-dimensional problem is examined in Section \ref{sec:hd}.  Throughout $\pi(\cdot)$ denotes a probability distribution, and $\pi(x)$ its density with respect to Lebesgue measure.

Since an early version of this work appeared online some contributions to the literature have been made that are worthy of mention.  A Markov kernel constructed as a state-dependent mixture is introduced in \cite{maire2018markov} and its properties are studied in some cases that are similar in spirit to the model problem of Section \ref{sec:hd}. An algorithm called \emph{Directional Metropolis--Hastings}, which encompasses a specific instance of the PDRWM, is introduced and studied in \cite{mallik2017directional}, and a modification of the same idea is used to develop the \emph{Hop} kernel within the \emph{Hug and Hop} algorithm of \cite{ludkin2019hug}. Kamatani considers an algorithm designed for the infinite-dimensional setting in \cite{kamatani2017ergodicity} of a similar design to that discussed in \cite{rudolf2015generalization} and studies the ergodicity properties.

\section{Markov Chains \& Geometric Ergodicity}
\label{sec:markov}
We will work on the measurable space $(\X,\B)$, so that each $X_t \in \X$ for a discrete-time Markov chain $\ch{X}$ with time-homogeneous transition kernel $P: \X \times \B \to [0,1]$, where $P(x,A) = \mathbb{P}[X_{i+1} \in A | X_i = x]$ and $P^n(x,A)$ is defined similarly for $X_{i+n}$.  All chains we consider will have invariant distribution $\pi(\cdot)$, and be both $\pi$-irreducible and aperiodic, meaning $\pi(\cdot)$ is the limiting distribution from $\pi$-almost any starting point \cite{roberts2004general}.  We use $|\cdot|$ to denote the Euclidean norm.

In Markov chain Monte Carlo the objective is to construct estimators of $\mathbb{E}_\pi[f]$, for some $f:\X \to \mathbb{R}$, by computing
\[
\hat{f}_n = \frac{1}{n}\sum_{i=1}^n f(X_i), ~~ X_i \sim P^i(x_0,\cdot).
\]
If $\pi(\cdot)$ is the limiting distribution for the chain then $P$ will be \emph{ergodic}, meaning $\hat{f}_n \xrightarrow{a.s.} \mathbb{E}_\pi[f]$ from $\pi$-almost any starting point.  For finite $n$ the quality of $\hat{f}_n$ intuitively depends on how quickly $P^n(x,\cdot)$ approaches $\pi(\cdot)$.  We call the chain \emph{geometrically ergodic} if
\begin{equation} \label{eqn:ge}
\| P^n(x,\cdot) - \pi(\cdot) \|_{TV} \leq M(x)\rho^n,
\end{equation}
from $\pi$-almost any $x \in \X$, for some $M > 0$ and $\rho < 1$, where $\| \mu(\cdot) - \nu(\cdot) \|_{TV} :=  \sup_{A \in \B} |\mu(A) - \nu(B)|$ is the total variation distance between distributions $\mu(\cdot)$ and $\nu(\cdot)$ \cite{roberts2004general}.

For $\pi$-reversible Markov chains geometric ergodicity implies that if $\mathbb{E}_\pi[f^2] < \infty$ for some $f:\mathcal{X} \to \mathbb{R}$, then
\begin{equation} \label{eqn:mcclt}
\sqrt{n} \left( \hat{f}_n - \mathbb{E}_\pi[f] \right) \xrightarrow{d} \mathcal{N}\left( 0, v(P,f) \right),
\end{equation}
for some asymptotic variance $v(P,f)$ \cite{roberts1997geometric}.  Equation (\ref{eqn:mcclt}) enables the construction of asymptotic confidence intervals for $\hat{f}_n$.

In practice, geometric ergodicity does not guarantee that $\hat{f}_n$ will be a sensible estimator, as $M(x)$ can be arbitrarily large if the chain is initialised far from the typical set under $\pi(\cdot)$, and $\rho$ may be very close to 1.  However, chains which are not geometrically ergodic can often either get `stuck' for a long time in low-probability regions or fail to explore the entire distribution adequately, sometimes in ways that are difficult to diagnose using standard MCMC diagnostics.

\subsection{Establishing geometric ergodicity}

It is shown in Chapter 15 of \cite{meyn2009markov} that (\ref{eqn:ge}) is equivalent to the condition that there exists a \emph{Lyapunov} function $V:\X \to [1,\infty)$ and some $\lambda < 1,b < \infty$ such that
\begin{equation} \label{eqn:drift1}
PV(x) \leq \lambda V(x) + b\mathbb{I}_C(x),
\end{equation}
where $PV(x):=\int V(y)P(x,dy)$.  The set $C \subset \X$ must be \emph{small}, meaning that for some $m \in \mathbb{N}$, $\varepsilon > 0$ and probability measure $\nu(\cdot)$
\begin{equation} \label{eqn:minor}
P^m(x,A) \geq \varepsilon \nu(A),
\end{equation}
for any $x \in C$ and $A \in \B$.  Equations (\ref{eqn:drift1}) and (\ref{eqn:minor}) are referred to as \emph{drift} and \emph{minorisation} conditions.  Intuitively, $C$ can be thought of as the centre of the space, and (\ref{eqn:drift1}) ensures that some one dimensional projection of $\ch{X}$ drifts towards $C$ at a geometric rate when outside.  In fact, (\ref{eqn:drift1}) is sufficient for the return time distribution to $C$ to have geometric tails \cite{meyn2009markov}.  Once in $C$, (\ref{eqn:minor}) ensures that with some probability the chain forgets its past and hence \emph{regenerates}.  This regeneration allows the chain to couple with another initialised from $\pi(\cdot)$, giving a bound on the total variation distance through the \emph{coupling inequality} (e.g. \cite{roberts2004general}).  More intuition is given in \cite{jones2001honest}.

Transition kernels considered here will be of the \emph{Metropolis--Hastings} type, given by
\begin{equation} \label{eqn:mh}
P(x,dy) = \alpha(x,y)Q(x,dy) + r(x) \delta_x(dy),
\end{equation}
where $Q(x,dy) = q(y|x)dy$ is some candidate kernel, $\alpha$ is called the acceptance rate and $r(x) = 1 - \int \alpha(x,y)Q(x,dy)$. Here we choose
\begin{equation} \label{eqn:arate}
\alpha(x,y) = 1 \wedge \frac{\pi(y)q(x|y)}{\pi(x)q(y|x)},
\end{equation}
where $a \wedge b$ denotes the minimum of $a$ and $b$.  This choice implies that $P$ satisfies detailed balance for $\pi(\cdot)$ \cite{tierney1994markov}, and hence the chain is $\pi$-reversible (note that other choices for $\alpha$ can result in non-reversible chains, see \cite{bierkens2014non} for details).

Roberts \& Tweedie \cite{roberts1996geometric}, following on from \cite{meyn2009markov}, introduced the following regularity conditions.

\begin{Theorem} \label{thm:roberts} 
(Roberts \& Tweedie).  Suppose that $\pi(x)$ is bounded away from $0$ and $\infty$ on compact sets, and there exists $\delta_q > 0$ and $\varepsilon_q >0$ such that, for every $x$
\[
|x-y| \leq \delta_q \Rightarrow q(y|x) \geq \varepsilon_q.
\]
Then the chain with kernel (\ref{eqn:mh}) is $\mu^{Leb}$-irreducible and aperiodic, and every nonempty compact set is small.
\label{thm:reg}
\end{Theorem}

For the choices of $Q$ considered in this article these conditions hold, and we will restrict ourselves to forms of $\pi(x)$ for which the same is true (apart from a specific case in Section \ref{sec:hd}).  Under Theorem \ref{thm:reg} then (\ref{eqn:ge}) only holds if a Lyapunov function $V:\X \to [1,\infty]$ with $\mathbb{E}_\pi[V] < \infty$ exists such that
\begin{equation} \label{eqn:driftcond}
\limsup_{|x| \to \infty} \frac{PV(x)}{V(x)} < 1.
\end{equation}
When $P$ is of the Metropolis-Hastings type, (\ref{eqn:driftcond}) can be written
\begin{equation} \label{eqn:dc2}
\limsup_{|x| \to \infty} \int \left[ \frac{V(y)}{V(x)} - 1 \right] \alpha(x,y)Q(x,dy) < 0.
\end{equation}
In this case a simple criterion for lack of geometric ergodicity is
\begin{equation} \label{eqn:lackge}
\limsup_{|x| \to \infty} r(x) = 1.
\end{equation}
Intuitively this implies that the chain is likely to get `stuck' in the tails of a distribution for large periods.

Jarner \& Tweedie \cite{jarner2003necessary} introduce a necessary condition for geometric ergodicity through a \emph{tightness} condition.

\begin{Theorem} \label{thm:jarner}
(Jarner \& Tweedie).  If for any $\varepsilon > 0$ there is a $\delta > 0$ such that for all $x \in \X$
\[
P(x,B_\delta(x)) > 1 - \varepsilon,
\]
where $B_\delta(x) := \{ y \in \X : d(x,y) < \delta \}$, then a necessary condition for $P$ to produce a geometrically ergodic chain is that for some $s > 0$
\[
\int e^{s|x|} \pi(dx) < \infty.
\]
\end{Theorem} 
The result highlights that when $\pi(\cdot)$ is heavy-tailed the chain must be able to make very large moves and still be capable of returning to the centre quickly for (\ref{eqn:ge}) to hold.

\section{Position-dependent Random Walk Metropolis}
\label{sec:pdrwm}
In the RWM, $Q(x,dy) = q(y-x)dy$ with $q(y-x)=q(x-y)$, meaning (\ref{eqn:arate}) reduces to $\alpha(x,y) = 1 \wedge \pi(y)/\pi(x)$.  A common choice is $Q(x,\cdot) = \mathcal{N}(x,h\Sigma)$, with $\Sigma$ chosen to mimic the global covariance structure of $\pi(\cdot)$ \cite{sherlock2010random}.  Various results exist concerning the optimal choice of $h$ in a given setting (e.g. \cite{roberts2001optimal}).  It is straightforward to see that Theorem \ref{thm:jarner} holds here, so that the tails of $\pi(x)$ must be uniformly exponential or lighter for geometric ergodicity.  In one dimension this is in fact a sufficient condition \cite{mengersen1996rates}, while for higher dimensions additional conditions are required \cite{roberts1996geometric}.  We return to this case in Section \ref{sec:hd}.

In the PDRWM $Q(x,\cdot) = \mathcal{N}(x,hG(x)^{-1})$, so (\ref{eqn:arate}) becomes
\[
\alpha(x,y) = 1 \wedge\frac{\pi(y)|G(y)|^{\frac{1}{2}}}{\pi(x)|G(x)|^{\frac{1}{2}}}\exp \left( -\frac{1}{2} (x -y)^T [G(y) - G(x)] (x-y) \right).
\]
%Note that throughout $G^{-1}(x) = [G(x)]^{-1}$, the inverse of the matrix $G(x)$, rather than the inverse function $G^{-1}$ such that $G \circ G^{-1}(x) = x$. There is no need for the function $G(x)$ to be invertible.
The motivation for designing such an algorithm is that proposals are more able to reflect the local dependence structure of $\pi(\cdot)$.  In some cases this dependence may vary greatly in different parts of the state-space, making a global choice of $\Sigma$ ineffective \cite{sejdinovic2014kernel}.

Readers familiar with differential geometry will recognise the volume element $|G(x)|^{1/2}dx$ and the linear approximations to the distance between $x$ and $y$ taken at each point through $G(x)$ and $G(y)$ if $\X$ is viewed as a Riemannian manifold with metric $G$.  We do not explore these observations further here, but the interested reader is referred to \cite{livingstone2014information} for more discussion.

The choice of $G(x)$ is an obvious question.  In fact, specific variants of this method have appeared on many occasions in the literature, some of which we now summarise.

\begin{enumerate}
\item \emph{Tempered Langevin diffusions} \cite{roberts2002langevin} $G(x) = \pi(x)I$.  The authors highlight that the diffusion with dynamics $dX_t = \pi^{-\frac{1}{2}}(X_t)dW_t$ has invariant distribution $\pi(\cdot)$, motivating the choice.  The method was shown to perform well for a bi-modal $\pi(x)$, as larger jumps are proposed in the low density region between the two modes.
\item \emph{State-dependent Metropolis} \cite{roberts2009examples} $G(x) = (1+|x|)^{-b}$.  Here the intuition is simply that $b>0$ means larger jumps will be made in the tails.  In one dimension the authors compare the expected squared jumping distance $\mathbb{E}[(X_{i+1} - X_i)^2]$ empirically for chains exploring a $\mathcal{N}(0,1)$ target distribution, choosing $b$ adaptively, and found $b \approx 1.6$ to be optimal.
\item \emph{Regional adaptive Metropolis--Hastings} \cite{roberts2009examples,craiu2009learn}. $G(x)^{-1} = \sum_{i=1}^m \mathbb{I}(x \in \X_i)\Sigma_i$.  In this case the state-space is partitioned into $\X_1 \cup ... \cup \X_m$, and a different proposal covariance $\Sigma_i$ is learned adaptively in each region $1 \leq i \leq m$.  An extension which allows for some errors in choosing an appropriate partition is discussed in \cite{craiu2009learn}
\item \emph{Localised Random Walk Metropolis} \cite{andrieu2008tutorial}. $G(x)^{-1} = \sum_{k=1}^m \check{q}_{\theta}(k|x) \Sigma_k$.  Here $\check{q}_{\theta}(k|x)$ are weights based on approximating $\pi(x)$ with some mixture of Normal/Student's t distributions, using the approach suggested in \cite{andrieu2006ergodicity}.  At each iteration of the algorithm a mixture component $k$ is sampled from $\check{q}_{\theta}(\cdot|x)$, and the covariance $\Sigma_k$ is used for the proposal $Q(x,dy)$.
\item \emph{Kernel adaptive Metropolis--Hastings} \cite{sejdinovic2014kernel}. $G(x)^{-1} = \gamma^2 I + \nu^2 M_x H M_x^T$, where $M_x = 2[\nabla_x k(z_1,x),\allowbreak...,\nabla_x k(z_n,x)]$ for some kernel function $k$ and $n$ past samples $\{z_1,...,z_n\}$, $H = I - (1/n) \mathbf{1}_{n\times n}$ is a centering matrix (the $n \times n$ matrix $\mathbf{1}_{n\times n}$ has 1 as each element), and $\gamma$, $\nu$ are tuning parameters.  The approach is based around performing nonlinear principal components analysis on past samples from the chain to learn a local covariance.  Illustrative examples for the case of a Gaussian kernel show that $M_xHM_x^T$ acts as a weighted empirical covariance of samples $z$, with larger weights given to the $z_i$ which are closer to $x$ \cite{sejdinovic2014kernel}.
\end{enumerate}
The latter cases also motivate any choice of the form
\[
G(x)^{-1} = \sum_{i=1}^n w(x,z_i) (z_i - x)^T(z_i-x)
\]
for some past samples $\{z_1,...,z_n\}$ and weight function $w:\X \times \X \to [0,\infty)$ with $\sum_i w(x,z_i) = 1$ that decays as $|x - z_i|$ grows, which would also mimic the local curvature of $\pi(\cdot)$ (taking care to appropriately regularise and diminish adaptation so as to preserve ergodicity, as outlined in \cite{andrieu2008tutorial}).

Some of the above schemes are examples of adaptive MCMC, in which a candidate from among a family of Markov kernels $\{P_\theta : \theta \in \Theta\}$ is selected by learning the parameter $\theta \in \Theta$ during the simulation \cite{andrieu2008tutorial}.  Additional conditions on the adaptation process (i.e. the manner in which $\theta$ is learned) are required to establish ergodicity results for the resulting stochastic processes.  We consider the decisions on how to learn $\theta$ appropriately to be a separate problem and beyond the scope of the present work, and instead focus attention on establishing geometric ergodicity of the base kernels $P_\theta$ for any fixed $\theta \in \Theta$. We note that this is typically a pre-requisite for establishing convergence properties of any adaptive MCMC method \cite{andrieu2008tutorial}.

\section{Results in one dimension}
\label{sec:1d}

Here we consider two different general scenarios as $|x| \to \infty$, i) $G(x)$ is bounded above and below, and ii) $G(x) \to 0$ at some specified rate.  Of course there is also the possibility that $G(x) \to \infty$, though intuitively this would in chains that spend a long time in the tails of a distribution, so we do not consider it (if $G(x) \to \infty$ then chains will in fact exhibit the \emph{negligible moves} property studied in \cite{livingstone2019kinetic}).

We begin with a result to emphasise that a growing variance as a necessary requirement for geometric ergodicity in the heavy-tailed case. 

\begin{Proposition} \label{prop:1}
If $G(x) \geq \sigma^{-2}$ for some $\sigma^{-2}>0$, then unless $\int e^{\eta|x|}\pi(dx) < \infty$ for some $\eta>0$ the PDRWM cannot produce a geometrically ergodic Markov chain.
\end{Proposition}

The above is a simple extension of a result that is well-known in the RWM case.  Essentially the tails of the distribution should be exponential or lighter to ensure fast convergence.  This motivates consideration of three different types of behaviour for the tails of $\pi(\cdot)$.

\begin{Assumption}
The density $\pi(x)$ satisfies one of the following tail conditions for all $y,x \in \mathcal{X}$ such that $|y| > |x| > t$, for some finite $t >0$.
\begin{enumerate}
    \item $\pi(y)/\pi(x) \leq \exp \{-a(|y|-|x|) \}$ for some $a>0$
    \item $\pi(y)/\pi(x) \leq \exp \{ -a(|y|^\beta-|x|^\beta) \}$ for some $a>0$ and $\beta \in (0,1)$
    \item $\pi(y)/\pi(x) \leq \left(|x|/|y|\right)^p$ for some $p > 1$.
\end{enumerate}
\end{Assumption}

Naturally Assumption 1.1 implies 1.2 and 1.2 implies 1.3. If Assumption 1.1 is not satisfied then $\pi(\cdot)$ is generally called \emph{heavy-tailed}.  When $\pi(x)$ satisfies Assumption 1.2 or 1.3 but not 1.1, then the RWM typically fails to produce a geometrically ergodic chain \cite{mengersen1996rates}.  We show in the sequel, however, that this is not always the case for the PDRWM.  We assume the below assumptions for $G(x)$ to hold throughout this section.

\begin{Assumption}
The function $G: \mathcal{X} \to (0,\infty)$ is bounded above by some $\sigma_b^{-2} < \infty$, and bounded below for all $x \in \mathcal{X}$ with $|x| < t$, for some $t > 0$. 
\end{Assumption}

The heavy-tailed case is known to be a challenging scenario, but the RWM will produce a geometrically ergodic Markov chain if $\pi(x)$ is log-concave.  Next we extend this result to the case of sub-quadratic variance growth in the tails.

\begin{Proposition} \label{prop:2}
If $\exists r<\infty$ such that $G(x) \propto |x|^{-\gamma}$ whenever $|x| > r$, then the PDRWM will produce a geometrically ergodic chain in both of the following cases:
\begin{enumerate}
    \item $\pi(x)$ satisfies Assumption 1.1 and $\gamma \in [0,2)$
    \item $\pi(x)$ satisfies Assumption 1.2 for some $\beta \in (0,1)$ and $\gamma  \in (2(1-\beta),2)$
\end{enumerate}
\end{Proposition}

The second part of Proposition \ref{prop:2} is not true for the RWM, for which Assumption 1.2 alone is not sufficient for geometric ergodicity \citep{mengersen1996rates}.

We do not provide a complete proof that the PDRWM will not produce a geometrically ergodic chain when only Assumption 1.3 holds and $G(x) \propto |x|^{-\gamma}$ for some $\gamma<2$, but do show informally that this will be the case.  Assuming that in the tails $\pi(x) \propto |x|^{-p}$ for some $p>1$ then for large $x$
\begin{equation} \label{eqn:rwtails}
\alpha(x,x+cx^{\gamma/2}) = 1\wedge \left(\frac{x}{x + cx^{\gamma/2}}\right)^{p+\gamma/2}\exp \left( - \frac{c^2x^\gamma}{2h}\left[ \frac{1}{(x+cx^{\gamma/2})^\gamma} - \frac{1}{x^\gamma} \right] \right).
\end{equation}
The first expression on the right hand side converges to $1$ as $x \to \infty$, which is akin to the case of fixed proposal covariance.  The second term will be larger than one for $c>0$ and less than one for $c<0$.  So the algorithm will exhibit the same `random walk in the tails' behaviour which is often characteristic of the RWM in this scenario, meaning that the acceptance rate fails to enforce a geometric drift back into the centre of the space.

When $\gamma = 2$ the above intuition will not necessarily hold, as the terms in \eqref{eqn:rwtails} will be roughly constant with $x$.  When only Assumption 1.3 holds it is therefore tempting to make the choice $G(x) = x^{-2}$ for $|x| >r$.  Informally we can see that such behaviour may lead to a favourable algorithm if a small enough $h$ is chosen. For any fixed $x>r$ a typical proposal will now take the form $y = (1+\xi\sqrt{h})x$, where $\xi \sim N(0,1)$. It therefore holds that
\begin{equation} \label{eqn:mrwm}
y = e^{\xi\sqrt{h}}x + O(\xi^2h x).
\end{equation}
The first term on the right-hand side of \eqref{eqn:mrwm} corresponds to the proposal of the \emph{multiplicative Random Walk Metropolis}, which is known to be geometrically ergodic under Assumption 1.3 (e.g. \cite{sherlock2010random}), as this equates to taking a logarithmic transformation of $x$, which `lightens' the tails of the target density to the point where it becomes log-concave. So in practice we can expect good performance from this choice of $G(x)$.  The above intuition does not, however, provide enough to establish geometric ergodicity, as the final term on the right-hand side of \eqref{eqn:mrwm} grows unboundedly with $x$ for any fixed choice of $h$.  The difference between the acceptance rates of the multiplicative Random Walk Metropolis and the PDRWM with $G(x) = x^{-2}$ will be the exponential term in \eqref{eqn:rwtails}. This will instead become polynomial by letting the proposal noise $\xi$ follow a distribution with polynomial tails (e.g. student's t), which is known to be a favourable strategy for the RWM when only Assumption 1.3 holds \cite{jarner2007convergence}. One can see that if the heaviness of the proposal distribution is carefully chosen then the acceptance rate may well enforce a geometric drift into the centre of the space, though for brevity we restrict attention to Gaussian proposals in this article.

The final result of this section provides a note of warning that lack of care in choosing $G(x)$ can have severe consequences for the method.

\begin{Proposition} \label{prop:4}
If $G(x)x^2 \to 0$ as $|x| \to \infty$, then the PDRWM will not produce a geometrically ergodic Markov chain.
\end{Proposition}

The intuition for this result is straightforward when explained.  In the tails, typically $|y-x|$ will be the same order of magnitude as $\sqrt{G(x)^{-1}}$, meaning $|y-x|/|x|$ grows arbitrarily large as $|x|$ grows.  As such, proposals will `overshoot' the typical set of the distribution, sending the sampler further out into the tails, and will therefore almost always be rejected.  The result can be related superficially to a lack of geometric ergodicity for Metropolis--Hastings algorithms in which the proposal mean is comprised of the current state translated by a drift function (often based in $\nabla\log\pi(x)$) when this drift function grows faster then linearly with $|x|$ (e.g. \cite{roberts1996exponential,livingstone2019geometric}).

\section{A higher-dimensional case study}
\label{sec:hd}

An easy criticism of the above analysis is that the one-dimensional scenario is sometimes not indicative of the more general behaviour of a method.  We note, however, that typically the geometric convergence properties of Metropolis--Hastings algorithms do carry over somewhat naturally to more than one dimension when $\pi(\cdot)$ is suitably regular (e.g. \cite{roberts1996geometric,jarner2000geometric}). Because of this we expect that the growth conditions specified above could be supplanted onto the determinant of $G(x)$ when the dimension is greater than one (leaving the details of this argument for future work).

% The following result carries over the the higher-dimensional setting without modification.
%\begin{Proposition} \label{prop:5}
%If each element of $G^{-1}(x)$ is bounded above (uniformly in $x$), then it is necessary that $\int e^{\eta|x|} \pi(dx) < \infty$ for some $\eta > 0$ for the PDRWM to produce a geometrically ergodic Markov chain when targetting $\pi(\cdot)$.
%\end{Proposition}

A key difference in the higher-dimensional setting is that $G(x)$ now dictates both the \emph{size} and \emph{direction} of proposals.  In the case $G(x)^{-1} = \Sigma$, some additional regularity conditions on $\pi(x)$ are required for geometric ergodicity in more than one dimension, outlined in  \cite{roberts1996geometric,jarner2000geometric}.  An example is also given in \cite{roberts1996geometric} of the simple two-dimensional density $\pi(x,y) \propto \exp( -x^2 - y^2 - x^2 y^2 )$, which fails to meet these criteria.  The difficult models are those for which probability concentrates on a ridge in the tails, which becomes ever narrower as $|x|$ increases.  In this instance, proposals from the RWM are less and less likely to be accepted as $|x|$ grows.  Another well-known example of this phenomenon is the \emph{funnel} distribution introduced in \cite{neal2003slice}.  

%\begin{figure}[ht]
%\begin{center}
%\includegraphics[width = 10cm,height = 7cm,trim=2cm 2cm 2cm 0cm]{Figures/2Dplot_is.pdf}
%\end{center}
%\label{fig:2d}
%\caption{Contours of the density $\pi(x,y) \propto \exp( - x^2 - y^2 - x^2 y^2 )$.  The left-hand plots show that a RWM with spherical covariance will find it increasingly difficult to propose values which will be accepted as the chain moves into the tails.  The right-hand plots suggest that allowing the covariance to change with position might alleviate this issue.}
%\end{figure}

\begin{figure}[ht]
\begin{center}
\includegraphics[height = 7cm]{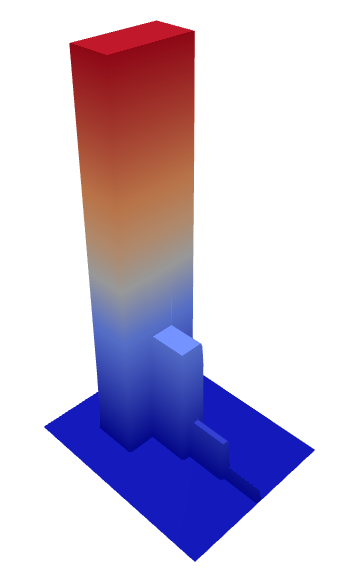}
\end{center}
\caption{The staircase distribution, with density given by \eqref{eqn:staircase}.}
\label{fig:rec1}
\end{figure}

To explore the behaviour of the PDRWM in this setting, we design a model problem, the \emph{staircase} distribution, with density
\begin{equation} \label{eqn:staircase}
\mathfrak{s}(x) \propto 3^{-\lfloor x_2 \rfloor}\mathbb{I}_R(x), ~~ R := \{ y \in \mathbb{R}^2 ; y_2 \geq 1, |y_1| \leq 3^{1-\lfloor y_2 \rfloor} \},
\end{equation}
where $\lfloor z \rfloor$ denotes the integer part of $z >0$.  Graphically the density is a sequence of cuboids on the upper-half plane of $\mathbb{R}^2$ (starting at $y_2 = 1$), each centred on the vertical axis, with height one and with each successive cuboid one third of the width and depth of the previous.  The density resembles an ever narrowing staircase, as shown in Figure \ref{fig:rec1}.

We denote by $Q_R$ the proposal kernel associated with the Random Walk Metropolis algorithm with fixed covariance $h\Sigma$. In fact the specific choice of $h$ and $\Sigma$ does not matter provided that the result is positive-definite.  For the PDRWM we denote by $Q_P$ the proposal kernel with covariance matrix
$$
hG(x)^{-1} = 
\left(
\begin{array}{cc}
3^{- 2 \lfloor x_2 \rfloor} & 0 \\
0 & 1
\end{array}
\right),
$$
which will naturally adapt the scale of the first coordinate to the width of the ridge.

\begin{Proposition} \label{prop:2dRWM}
The Metropolis--Hastings algorithm with proposal $Q_R$ does not produce a geometrically ergodic Markov chain when $\pi(x) = \mathfrak{s}(x)$.
\end{Proposition}

The design of the PDRWM proposal kernel $Q_P$ in this instance is such that the proposal covariance reduces at the same rate as the width of the stairs, therefore naturally adapting the proposal to the width of the ridge on which the density concentrates. This state-dependent adaptation results in a geometrically ergodic chain, as shown in the below result.

\begin{Proposition} \label{prop:2dPDRWM}
The Metropolis--Hastings algorithm with proposal $Q_P$ produces a geometrically ergodic Markov chain when $\pi(x) = \mathfrak{s}(x)$.
\end{Proposition}

\section{Discussion}
\label{sec:discussion}

In this paper we have analysed the ergodic behaviour of a Metropolis-Hastings method with proposal kernel $Q(x,\cdot) = \mathcal{N}(x,hG(x)^{-1})$.  In one dimension we have characterised the behaviour in terms of growth conditions on $G(x)^{-1}$ and tail conditions on the target distribution, and in higher dimensions a carefully constructed model problem is discussed.  The fundamental question of interest was whether generalising an existing Metropolis--Hastings method by allowing the proposal covariance to change with position can alter the ergodicity properties of the sampler.  We can confirm that this is indeed possible, either for the better or worse, depending on the choice of covariance.  The take home points for practitioners are i) lack of sufficient care in the design of $G(x)$ can have severe consequences (as in Proposition \ref{prop:4}), and ii) careful choice of $G(x)$ can have much more beneficial ones, perhaps the most surprising of which are in the higher-dimensional setting, as evidenced in Section \ref{sec:hd}.  

We feel that such results can also offer insight into similar generalisations of different Metropolis--Hastings algorithms (e.g. \cite{girolami2011riemann,xifara2014langevin}).  For example, it seems intuitive that any method in which the variance grows at a faster than quadratic rate in the tails is unlikely to produce a geometrically ergodic chain.  There are connections between the PDRWM and some extensions of the Metropolis-adjusted Langevin algorithm \cite{xifara2014langevin}, the ergodicity properties of which are discussed in \cite{latuszynski2011discussion}.  The key difference between the schemes is the inclusion of the drift term $G(x)^{-1}\nabla\log\pi(x)/2$ in the latter.  It is this term which in the main governs the behaviour of the sampler, which is why the behaviour of the PDRWM is different to this scheme.

We can apply these results to the specific variants discussed in Section 3.  Provided sensible choices of regions/weights, and diminishing adaptation schemes are chosen, the Regional adaptive Metropolis--Hastings, Locally weighted Metropolis and Kernel-adaptive Metropolis--Hastings samplers should all satisfy $G(x)\to\Sigma$ as $|x| \to \infty$, meaning they can be expected to inherit the ergodicity properties of the standard RWM (the behaviour in the centre of the space, however, will likely be different).  In the State-dependent Metropolis method provided $b<2$ then the sampler should also behave reasonably.  Whether or not a large enough value of $b$ would be found by a particular adaptation rule is not entirely clear, and this could be an interesting direction of further study.  The Tempered Langevin diffusion scheme, however, will fail to produce a geometrically ergodic Markov chain whenever the tails of $\pi(x)$ are lighter than that of a Cauchy distribution.  To allow reasonable tail exploration when this is the case, two pragmatic options would be to upper bound $G(x)^{-1}$ manually or use this scheme in conjunction with another, as there is evidence that the sampler can perform favourably when exploring the centre of a distribution \cite{roberts2002langevin}.  None of the specific variants discussed here are able to mimic the local curvature of the $\pi(x)$ in the tails, so as to enjoy the favourable behaviour exemplified in Proposition \ref{prop:2dPDRWM}.  This is possible using Hessian information as in \cite{girolami2011riemann}, but should also be possible in some cases using appropriate surrogates.

\section*{Acknowledgments}
This research was supported by a UCL IMPACT PhD scholarship co-funded by Xerox Research Centre Europe and EPSRC.  The author thanks Alexandros Beskos, Krzysztof \L atuszy\'{n}ski and Gareth Roberts for several useful discussions, Michael Betancourt for proofreading the paper, and Mark Girolami for general supervision and guidance.

%% Appendices
\appendix
\section{Proofs}

%%%%%%%%%%%%%%%%%%%%%%%%%%%%%%%%%%%
%%%%%%%%%%%%%%%%%%%%%%%%%%%%%%%%%%%
%%%%%%%%% 1D Proposition 1    %%%%%
%%%%%%%%%%%%%%%%%%%%%%%%%%%%%%%%%%%
%%%%%%%%%%%%%%%%%%%%%%%%%%%%%%%%%%%

\begin{proof}[Proof of Proposition \ref{prop:1}]
In this case for any choice of $\varepsilon>0$ there is a $\delta > 0$ such that $Q(x,B_\delta(x)) > 1 - \varepsilon$. Noting that $P(x,B_\delta(x)) \geq Q(x,B_\delta(x))$ when $P$ is of Metropolis--Hastings type, Theorem 2 can be applied directly.
\end{proof}

%%%%%%%%%%%%%%%%%%%%%%%%%%%%%%%%%%%
%%%%%%%%%%%%%%%%%%%%%%%%%%%%%%%%%%%
%%%%%%%%% 1D Proposition 2    %%%%%
%%%%%%%%%%%%%%%%%%%%%%%%%%%%%%%%%%%
%%%%%%%%%%%%%%%%%%%%%%%%%%%%%%%%%%%

\begin{proof}[Proof of Proposition \ref{prop:2}]
For the log-concave case, take $V(x) = e^{s|x|}$ for some $s > 0$, and let $B_A$ denote the integral (\ref{eqn:dc2}) over the set $A$.  We first break up $\X$ into $(-\infty,0] \cup (0,x - cx^{\frac{\gamma}{2}}] \cup (x - cx^{\frac{\gamma}{2}}, x + cx^{\frac{\gamma}{2}}] \cup (x + cx^{\frac{\gamma}{2}}, x + cx^{\gamma}] \cup (x + cx^{\gamma}, \infty)$ for some $x>0$ and fixed constant $c \in (0,\infty)$, and show that the integral is strictly negative on at least one of these sets, and can be made arbitrarily small as $x \to \infty$ on all others.  The $-\infty$ case is analogous from the tail conditions on $\pi(x)$. From the conditions we can choose $x > r$ and therefore write $G(x)^{-1} = \eta x^\gamma$ for some fixed $\eta < \infty$.

On $(-\infty,0]$, we have
\begin{align*}
B_{(-\infty,0]} &= e^{-sx}\int_{-\infty}^0 e^{s|y|} \alpha(x,y)Q(x,dy) - \int_{-\infty}^0 \alpha(x,y)Q(x,dy) , \\
&\leq e^{-sx}\int_0^{\infty} e^{sy} Q(-x,dy).
\end{align*}
The integral is now proportional to the moment generating function of a truncated Gaussian distribution (see Appendix \ref{app:trunc}), so is given by
\[
e^{-sx+  h\eta x^\gamma s^2 / 2} \left[ 1 - \Phi \left( x^{1- \gamma / 2}/\sqrt{h\eta} - \sqrt{h\eta} sx^{\gamma/2} \right) \right].
\]
A simple bound on the error function is $\sqrt{2\pi}x\Phi^c(x) < e^{-x^2 / 2}$ \cite{johnDcook}, so setting $\vartheta = x^{1- \gamma / 2}/\sqrt{h\eta} - \sqrt{h\eta} sx^{\gamma/2}$ we have
\begin{align*}
B_{(-\infty,0]} &\leq \frac{1}{\sqrt{2\pi}} \exp \left(-2sx+ \frac{h\eta s^2}{2}x^{\gamma} -\frac{1}{2}\left( \frac{1}{h\eta}x^{2 - \gamma} - 2sx + h\eta s^2x^{\gamma} \right) + \log \vartheta \right), \\
&= \frac{1}{\sqrt{2\pi}} \exp \left(-sx -\frac{1}{2h\eta}x^{2 - \gamma} + \log \vartheta \right).
\end{align*}
which $\to 0$ as $x \to \infty$, so can be made arbitrarily small.

On $(0,x - cx^{\gamma/2}]$, note that $e^{s(|y| - |x|)} - 1$ is clearly negative throughout this region provided that $c < x^{1-\gamma/2}$, which can be enforced by choosing $x$ large enough for any given $c < \infty$.  So the integral is straightforwardly bounded as $B_{(0,x - cx^{\gamma/2}]}\leq 0$ for all $x \in \X$.

On $(x - cx^{\gamma/2},x+cx^{\gamma/2}]$, provided $x - cx^{\gamma/2} > r$ then for any $y$ in this region we can either upper or lower bound $\alpha(x,y)$ with the expression
\[
\exp\left( - a(y - x) + \frac{\gamma}{2}\log\left|\frac{x}{y}\right| -\frac{1}{2h\eta}\left[(x-y)^2 y^{-\gamma} - (x-y)^2 x^{-\gamma} \right] \right).
\]
A Taylor expansion of $y^{-\gamma}$ about $x$ gives
\[
y^{-\gamma} = x^{-\gamma} - \gamma x^{-\gamma - 1}(y - x) + \gamma(\gamma + 1)x^{-\gamma - 2}(y-x)^2 + ...
\]
and multiplying by $(y-x)^2$ gives
\[
(y-x)^2 y^{-\gamma} = \frac{(y-x)^2}{x^\gamma} - \gamma\frac{(y-x)^3}{x^{\gamma + 1}} + \gamma(\gamma + 1)\frac{(y-x)^4}{x^{\gamma + 2}} + ...
\]
If $|y-x| = cx^{\gamma/2}$ then this is:
\[
\frac{c^2 x^{\gamma} }{x^\gamma} - \gamma\frac{c^3 x^{3\gamma/2}}{x^{\gamma + 1}} + \gamma(\gamma + 1)\frac{c^4 x^{2\gamma} }{x^{\gamma + 2}} + ...
\]
As $\gamma < 2$ then $3\gamma/2 < \gamma + 1$, and similarly for successive terms, meaning each gets smaller as $|x| \to \infty$.  So we have for large $x$, $y \in (x - cx^{\gamma/2},x + cx^{\gamma/2})$ and any $\delta>0$
\begin{equation} \label{eqn:lc2}
(y-x)^2 y^{-\gamma} \geq \frac{ (y-x)^2 }{ x^{\gamma} } - \gamma \frac{(y-x)^3}{x^{\gamma +1}} - 2h\eta\delta.
\end{equation}
%Using (\ref{eqn:lc2}) gives (for large enough $x$)
%\[
%\alpha(x,y) \leq \exp\left( -a (y- x) + \frac{\gamma}{2}\log\left|\frac{x}{y}\right| + \frac{1}{2h\eta}\gamma \frac{(y-x)^3}{x^{\gamma +1}} + \delta \right)
%\]
So we can analyse how the acceptance rate behaves.  First note that for fixed $\epsilon >0$
\[
\alpha(x,x+\epsilon) \leq \exp\left( -a \epsilon + \frac{\gamma}{2}\log\left|\frac{x}{x+\epsilon}\right| + \frac{1}{2h}\gamma \frac{\epsilon^3}{x^{\gamma +1}} + \delta \right) \to \exp( -a \epsilon +\delta),
\]
recalling that $\delta$ can be made arbitrarily small.  In fact it holds that the $e^{-a \epsilon}$ term will be dominant for any $\epsilon$ for which 
$\epsilon^3/x^{\gamma + 1} \to 0$, i.e. any  $\epsilon = o( x^{\gamma + 1/3})$.  If $\gamma < 2$ then $\epsilon = cx^{\gamma/2}$ satisfies this condition.  So for any $y > x$ in this region we can choose an $x$ such that
\[
\alpha(x,y) \leq \exp\left( -a (y - x) + \delta_x \right),
\]
where $\delta_x$ can be made arbitrarily small in this region by choosing a large enough $x$.  For the case $y < x$ here we have (for any fixed $\epsilon > 0$)
\[
\alpha(x,x-\epsilon) \geq \exp\left( a \epsilon + \frac{\gamma}{2}\log\left|\frac{x}{x-\epsilon}\right| - \frac{1}{2h}\gamma \frac{\epsilon^3}{x^{\gamma +1}} -\delta \right) \to \exp( a \epsilon -\delta).
\]
So by a similar argument we have $\alpha(x,y) > 1$ here when $x \to \infty$.  Combining gives
\begin{align*}
B_{(x - cx^{\gamma/2},x+cx^{\gamma/2}]} &\leq \int_0^{cx^{\gamma/2}} \left[ e^{ (s-a) z + \delta_z} - e^{-a z + \delta_z} + e^{-sz} - 1 \right] q_x(dz), \\
&= -\int_0^{cx^{\gamma/2}} (1 - e^{-sz})(1 - e^{(s-a)z + \delta_z} ) q_x(dz),
\end{align*}
which will be strictly negative for large enough $x$ provided $s < a$, where $q_x(\cdot)$ denotes a zero mean Gaussian distribution with the same variance as $Q(x,\cdot)$.

On $(x + cx^{\gamma/2}, x + cx^{\gamma}]$ we can upper bound the acceptance rate as
\[
\alpha(x,y) \leq \frac{\pi(y)}{\pi(x)}\exp \left( \frac{1}{2}\log\frac{|G(y)|}{|G(x)|} + \frac{G(x)}{2h}(x - y)^2 \right)
\]
If $y \geq x$ and $x > x_0$ then we have
\[
\alpha(x,y) \leq \exp \left( -a(|y| - |x|) + \frac{1}{2h\eta}\frac{(x-y)^2}{x^{\gamma}} \right).
\]
For $|y-x| = cx^{\ell}$ this becomes
\[
\alpha(x,y) \leq \exp \left( -a cx^{\ell} + \frac{c^2}{2h\eta} x^{2\ell - \gamma} \right)
\]
So provided $\gamma > \ell$ the first term inside the exponential will dominate the second for large enough $x$.  In the equality case we have
\[
\alpha(x,y) \leq \exp \left( \left(\frac{c^2}{2h\eta}-a \right) cx^{\gamma} \right),
\]
so provided we choose $c$ such that $a > c^2/(2h\eta)$ then the acceptance rate will also decay exponentially.  Because of this we have
\begin{align*}
B_{(x + cx^{\gamma/2}, x + cx^{\gamma}]} &\leq  \int_{x + cx^{\gamma/2}}^{x + cx^{\gamma}} e^{s(y - x)}\alpha(x,y)Q(x,dy), \\
&\leq e^{( c^2/(2h\eta) + s - a ) cx^{\gamma/2}}  Q(x,(x + cx^{\gamma/2}, x + cx^{\gamma}]),
\end{align*}
so provided $a > c^2/(2h\eta) + s$ then this term can be made arbitrarily small.

On $(x + cx^{\gamma}, \infty)$ using the same properties of truncated Gaussians we have
\begin{align*}
B_{(x + cx^{\gamma}, \infty)} &\leq e^{-sx}\int_{x + cx^{\gamma}}^\infty e^{sy}Q(x,dy), \\
&= e^{ s^2h\eta x^{\gamma}/2 }\Phi^c\left( \left(\frac{c}{\sqrt{h\eta}}-\sqrt{h\eta}s\right)x^{\gamma}\right),
\end{align*}
which can be made arbitrarily small provided that $s$ is chosen to be small enough using the same simple bound on $\Phi^c$ as for the case of $B_{(-\infty,0]}$.

Under Assumption 1.2 the proof is similar.  Take $V(x) = e^{s|x|^\beta}$, and divide $\X$ up into the same regions.  Outside of $(x-cx^{\gamma/2},x+cx^{\gamma/2}]$ the same arguments show that the integral can be made arbitrarily small.  On this set, note that in the tails
\[
(x + cx^\ell)^\beta - x^\beta = \beta c x^{\ell + \beta - 1} + \frac{\beta(\beta - 1)}{2}c^2x^{2\ell + \beta - 2} + ...
\]
For $y - x = cx^\ell$, then for $\ell < 1-\beta$ this becomes negligible.  So in this case we further divide the typical set into $(x, x+ cx^{1-\beta}] \cup (x + cx^{1-\beta},x + cx^{\gamma/2})$.  On $(x-cx^{1-\beta},x+cx^{1-\beta})$ the integral is bounded above by $e^{-c_1}Q(x,(x-cx^{1-\beta},x+cx^{1-\beta})) \to 0$, for some suitably chosen $c_1>0$.  On $(x - cx^{\gamma/2},x-cx^{1-\beta}] \cup (x + cx^{1-\beta},x + cx^{\gamma/2}]$ then for $y>x$ we have $\alpha(x,y) \leq e^{-c_2(y^\beta - x^\beta)}$, so we can use the same argument as in the the log-concave case to show that the integral will be strictly negative in the limit. 
\end{proof}

%%%%%%%%%%%%%%%%%%%%%%%%%%%%%%%%%%%
%%%%%%%%%%%%%%%%%%%%%%%%%%%%%%%%%%%
%%%%%%%%% 1D Proposition 4    %%%%%
%%%%%%%%%%%%%%%%%%%%%%%%%%%%%%%%%%%
%%%%%%%%%%%%%%%%%%%%%%%%%%%%%%%%%%%

\begin{proof}[Proof of Proposition \ref{prop:4}]
First note that in this case for any $g: \mathbb{R} \to (0,\infty)$ such that as $|x| \to \infty$ it holds that $g(x)/|x| \to \infty$ but $g(x)\sqrt{G(x)} \to 0$, then 
$$
Q(x,\{x-g(x),x+g(x)\}) = 
\Phi\left( g(x)\sqrt{G(x)} \right) 
- 
\Phi\left( -g(x)\sqrt{G(x)} \right) \to 0
$$
as $|x| \to \infty$.  The chain therefore has the property that $\mathbb{P}(\{|X_{i+1}| > g(X_i)/2 \} \cup \{ X_{i+1} = X_i \})$ can be made arbitrarily close to 1  as $|X_i|$ grows, which leads to two possible behaviours. If the form of $\pi(\cdot)$ enforces such large jumps to be rejected then $r(x) \to 1$ and lack of geometric ergodicity follows from \eqref{eqn:lackge}. If this is not the case then the chain will be transient (this can be made rigorous using a standard Borel--Cantelli argument, see e.g. the proof of Theorem 12.2.2 on p.299 of \cite{meyn2009markov}).
\end{proof}

%%%%%%%%%%%%%%%%%%%%%%%%%%%%%%%%%%%
%%%%%%%%%%%%%%%%%%%%%%%%%%%%%%%%%%%
%%%%%%%%% hd result Jarner %%%%%%%%
%%%%%%%%%%%%%%%%%%%%%%%%%%%%%%%%%%%
%%%%%%%%%%%%%%%%%%%%%%%%%%%%%%%%%%%

%\begin{proof}[Proof of Proposition \ref{prop:5}]
%As with Proposition \ref{prop:1}, a straightforward application of Theorem \ref{thm:jarner} gives the result.
%\end{proof}

\begin{proof}[Proof of Proposition \ref{prop:2dRWM}] %First note that although $\mathfrak{s}(x)$ is not bounded away from zero on compact sets \eqref{eqn:lackge}is still sufficient for lack of geometric ergodicity here. To see this note that for any fixed $m$ and $x$, for any choice of $\epsilon > 0$ there exists $\delta<\infty$ such that  $P^m(x,B_\delta(x)) >1-\epsilon$ where $B_\delta(x) := \{y \in \mathbb{R}^2 : |y-x| \leq \delta \}$.  If a set $C$ is small then there is a probability measure $\nu(\cdot)$ and a fixed constant $\varepsilon>0$ such that for all events $A$ in the Borel $\sigma$-field of $\mathbb{R}^2$
%\begin{equation} \label{eqn:smallset2}
%\inf_{x \in C} \frac{P^m(x,A)}{\nu(A)} > \varepsilon.
%\end{equation}
%For any fixed $\nu(\cdot)$ we can always choose a compact $A$ for which $\nu(A)$ is greater than zero (owing to the tightness of $\nu(\cdot)$).  If $C$ is unbounded, we can then choose $x$ such that $B_\delta(x) \cap A = \emptyset$, meaning the left-hand side of \eqref{eqn:smallset2} can be arbitrarily small, presenting a contradiction.

It is sufficient to construct a sequence of points $x_p \in \mathbb{R}^2$ such that $|x_p| \to \infty$ as $p \to \infty$, and show that $r(x_p) \to 1$ in the same limit, then apply \eqref{eqn:lackge}.  Take $x_p = (0,p)$ for $p \in \mathbb{N}$.  In this case
$$
r(x_p) = 1- \int \alpha(x_p,y)Q_R(x_p,dy)
$$
Note that for every $\epsilon>0$ there is a $\delta<\infty$ such that $Q(x_p, B_\delta^c(x_p))<\epsilon$ for all $x_p$, where $B_\delta(x) := \{y \in \mathbb{R}^2 : |y-x| \leq \delta \}$. The set $A(x_p,\delta) := B_\delta(x_p) \cap R$ 
denotes the possible values of $y \in B_\delta(x)$ for which the acceptance rate is non-zero. Note that $A(x_p,\delta) \subset S(x_p,\delta):= \{y \in B_\delta(x_p) : |y_1| \leq 3^{1-\lfloor p-\delta \rfloor} \}$, which is simply a strip that can be made arbitrarily narrow for any fixed $\delta$ by taking $p$ large enough.  Combining these ideas gives
$$
\begin{aligned}
\int \alpha(x_p,y)Q_R(x_p,dy) 
&\leq
\int_{A(x_p,\delta)} \alpha(x_p,y)Q_R(x_p,dy) + \epsilon \\
&\leq Q_R(x_p,S(x_p,\delta)) + \epsilon.
\end{aligned}
$$
Both of the quantities on the last line can be made arbitrarily small by choosing $p$ suitably large. Thus $r(x_p) \to 1$ as $|x_p| \to \infty$, as required.
\end{proof}

%%%%%%%%%%%%%%%%%%%%%%%%%%%%%%%%%%%
%%%%%%%%%%%%%%%%%%%%%%%%%%%%%%%%%%%
%%%%%%% 2D Result PDRWM %%%%%%%%%%%
%%%%%%%%%%%%%%%%%%%%%%%%%%%%%%%%%%%
%%%%%%%%%%%%%%%%%%%%%%%%%%%%%%%%%%%

\begin{proof}[Proof of Proposition \ref{prop:2dPDRWM}]  
First note that $\inf_{x \in R}Q_P(x,R)$ is bounded away from zero, unlike in the case of $Q_R$, owing to the design of $Q_P$. The acceptance rate here simplifies, since for any $y \in R$
$$
\frac{\mathfrak{s}(y)|G(y)|^{\frac{1}{2}}}{\mathfrak{s}(x)|G(x)|^{\frac{1}{2}}} = 1,
$$
meaning only the expression $\exp \left( -\frac{1}{2} (y-x)^T[G(y)-G(x)](y-x) \right)$ needs to be considered. In this case the expression is simply
$$
\exp\left( -\frac{1}{2} (3^{2\lfloor y_2 \rfloor} - 3^{2\lfloor x_2 \rfloor}) (y_1-x_1)^2 \right).
$$
Provided that $x_1 \neq y_1$, then when $1 \leq \lfloor y_2 \rfloor < \lfloor x_2 \rfloor$  this expression is strictly greater than 1, whereas in the reverse case it is strictly less than one. The resulting Metropolis--Hastings kernel $P$ using proposal kernel $Q_P$ will therefore satisfy $\int y_2 P(x,dy) < x_2$ for large enough $x_2$, and hence geometric ergodicity follows by taking the Lyapunov function $V(x) = e^{s|x_2|}$ (which can be used here since the domain of $x_1$ is compact) and following an identical argument to that given on pages 404-405 of \cite{meyn2009markov} for the case of the proof of geometric ergodicity of the random walk on the half-line model for suitably small $s > 0$, taking the small set $C := [0,1]\times [1,r]$ for suitably large $r < \infty$ and $\nu(\cdot) = \int_{\cdot}\mathfrak{s}(x)dx$.
\end{proof}

\section{Needed facts about truncated Gaussian distributions} \label{app:trunc}

Here we collect some elementary facts used in the article.  For more detail see e.g. \cite{johnson1970distributions}.  If $X$ follows a truncated Gaussian distribution $\mathcal{N}^T_{[a,b]}(\mu,\sigma^2)$ then it has density
\[
f(x) = \frac{1}{\sigma Z_{a,b}} \phi\left(\frac{x-\mu}{\sigma} \right)\mathbb{I}_{[a,b]}(x),
\]
where $\phi(x) = e^{-x^2/2}/\sqrt{2\pi}$, $\Phi(x) = \int_{-\infty}^x \phi(y)dy$ and $Z_{a,b} = \Phi( (b - \mu)/\sigma ) - \Phi( (a - \mu)/\sigma )$.  Defining $B = (b-\mu)/\sigma$ and $A = (a - \mu)/\sigma$, we have
\[
\mathbb{E}[X] = \mu + \frac{\phi(A) - \phi(B)}{Z_{a,b}}\sigma
\]
and
\[
\mathbb{E}[e^{tX}] = e^{\mu t + \sigma^2t^2/2} \left[ \frac{ \Phi(B - \sigma t) - \Phi(A - \sigma t)}{Z_{a,b}} \right].
\]
In the special case $b = \infty$, $a = 0$ this becomes $e^{\mu t + \sigma^2t^2/2}\Phi(\sigma t)/Z_{a,b}$.

%% Bibliography
\bibliographystyle{plain}
\bibliography{mybibfile}

\end{document}